%% file: ms.tex
\documentclass[12pt,preprint]{aastex}

\newcommand\UCHII{UCH\,{\sc ii}}
\newcommand\HII{H\,{\sc ii}}

\newcommand\kms{km~s$^{-1}$}

\newcommand\thCO{$^{13}$CO}
\newcommand\NHthree{NH$_3$}

\newcommand\cmthree{cm$^{-3}~$}
\newcommand\dycm{dyne cm$^{-2}~$}
\newcommand\etal{et al.~}
\newcommand\be{\begin{equation}}
\newcommand\ee{\end{equation}}
\newcommand\bea{\begin{eqnarray}}
\newcommand\eea{\end{eqnarray}}

\newcommand\ddeg{$^{o}$}

\catcode`\@=11
\newcommand{\gsim}{${\mathrel{\mathpalette\@versim>}}$}
\newcommand{\lsim}{${\mathrel{\mathpalette\@versim<}}$}
\newcommand{\@versim}[2]{\lower 2.9truept \vbox{\baselineskip 0pt \lineskip
    0.5truept \ialign{$\m@th#1\hfil##\hfil$\crcr#2\crcr\sim\crcr}}}
\catcode`\@=12

\slugcomment{To appear in ApJ}

\shorttitle{Properties of neutral material in the vicinity of an \UCHII\ region in W48}
\shortauthors{Roshi \etal }

\begin{document}

\title{Multi-wavelength carbon recombination line observations with the VLA toward an
       \UCHII\ region in W48: Physical properties and kinematics of neutral material}

\author{D. Anish Roshi}
\affil{Raman Research Institute, Sadashivanagar, Bangalore, India 560080, anish@rri.res.in  \\
and \\ National Radio Astronomy Observatory\altaffilmark{1}, Green Bank, USA}

\author{W. M. Goss}
\affil{National Radio Astronomy Observatory, P.O. Box O, Socorro, NM 87801, USA;
mgoss@nrao.edu}

\author{K. R. Anantharamaiah\altaffilmark{2}}
\affil{Raman Research Institute, Sadashivanagar, Bangalore, India 560080}

\and

\author{S. Jeyakumar\altaffilmark{3}} 
\affil{Raman Research Institute, Sadashivanagar, Bangalore, India 560080}

\altaffiltext{1}{The National Radio Astronomy Observatory is a facility of
the National Science Foundation operated under a cooperative
agreement by Associated Universities, Inc.}

\altaffiltext{2}{Deceased 29 October, 2001}

\altaffiltext{3}{Present address: Instituto de Geofisica, UNAM, Mexico 04521; 
sjk@soho.igeofcu.unam.mx }

\begin{abstract}

Using the Very Large Array (VLA) the C76$\alpha$ and C53$\alpha$ 
recombination lines (RLs) have been detected toward the ultra-compact \HII\
region (\UCHII\ region) G35.20$-$1.74. We also obtained 
upper limits to the carbon RLs at 6 cm (C110$\alpha$ \& C111$\alpha$) 
and 3.6 cm (C92$\alpha$) wavelengths with the VLA. In addition, continuum 
images of the W48A complex ( which includes G35.20$-$1.74 )
are made with angular resolutions in the range 14\arcsec\ to 2\arcsec.
Modeling the multi-wavelength line and continuum data has provided
the physical properties of the \UCHII\ region and the photodissociation
region (PDR) responsible for the carbon RL emission. 
The gas pressure in the PDR, estimated using the derived physical properties, 
is at least four times larger than that in the \UCHII\ region. 
The dominance of stimulated emission of carbon RLs 
near 2 cm, as implied by our models, is used to study the relative motion of
the PDR with respect to the molecular cloud and ionized gas.
Our results from the kinematical study are 
consistent with a pressure-confined \UCHII\ region with the ionizing
star moving with respect to the molecular cloud. 
However, based on the existing data, other models 
to explain the extended lifetime and morphology of \UCHII\ regions
cannot be ruled out.  

%In this paper, we show that multi-wavelength (0.7, 2, 3.6 \& 6 cm) 
%observations of carbon recombination lines (RLs) from the photo-dissociation
%region (PDR), formed in the dense molecular material surrounding
%\UCHII\ regions, can be used to study whether \UCHII\ regions are pressure
%confined. Pressure confinement of \UCHII\ regions has been 
%suggested as a possible explanation for the ``lifetime problem
%of \UCHII\ regions''. Here we present a study toward G35.20$-$1.74,
%a \UCHII\ region in the molecular cloud complex W48.
%Modeling the multi-wavelength line and continuum data has provided
%the physical properties of the \UCHII\ region and the PDR. These
%properties are used to estimate the gas pressure in the \HII\ region 
%and the PDR. The estimated gas pressure in the PDR is an order 
%of magnitude larger than that in the \UCHII\ region. The origin
%of this large pressure difference remains a puzzle and could not
%be understood with the existing data. Molecular line observations
%with similar angular resolution as that of RL data are needed
%to further investigate whether the PDR pressure represents the
%pressure of the ambient molecular cloud.
\end{abstract}


\section{Introduction}
\label{sec:intro}

Massive stars (OB stars) are formed by the gravitational
collapse of clumps in molecular clouds. Lyman continuum 
($E > 13.6$ eV) photons from newly born OB 
stars ionize the surrounding
material resulting in the formation of ultra-compact 
\HII\ regions (\UCHII\ regions). \UCHII\ regions are expected
to expand due to their high gas pressure. 
If they expand at the speed of sound in the ionized gas, then
the time taken for the \UCHII\ region to expand to $\sim$ 0.1 pc
is a few times  $10^3$ years. This time scale is referred to as
the dynamical lifetime. However, the lifetime
deduced from the observed number of \UCHII\ regions
is a few times  $10^5$ years.
%there is evidence that indicates that OB stars are embedded in their 
%parent molecular cloud for $\sim$ 15 \%
%of their lifetime.  During this phase the star will be associated
%with an \UCHII\ region implying that the lifetime of
%\UCHII\ regions is a few times  $10^5$ years.
This discrepancy between the two lifetimes is referred to
as the ``lifetime problem'' of \UCHII\ regions
(\nocite{wc89a}Wood \& Churchwell 1989a).
Resolving the lifetime problem is an important step
in understanding massive star formation in the Galaxy 
(see review by \nocite{gl99}Garay \& Lizano 1999).

There have been several suggestions to resolve the lifetime problem.
In particular, \nocite{drg95}De Pree, Rodr\'iguez \& Goss (1995)
proposed that if high density ($\sim$ 10$^7$ \cmthree), 
warm ($\sim$ 100 K) molecular material is present 
in the vicinity of \UCHII\ regions, it may be able to 
pressure confine \UCHII\ regions 
that form there and thus extend their lifetime. 
In the recent past, attempts have been made to 
observationally determine the physical properties
of the molecular material near \UCHII\ regions.
For example, \nocite{ac96}Akeson \& Carlstrom (1996) 
have used methyl cyanide, a good tracer of
the temperature and density of dense molecular
cores, to estimate the physical properties of the
ambient medium near \UCHII\ regions G5.89+0.4
and G34.3+0.2. They concluded that the estimated
ambient pressure ($>$ 10$^8$ K \cmthree) was high enough to 
pressure confine the \UCHII\ 
regions. The dominant component of the
ambient pressure need not always be the thermal pressure
of the molecular gas but can be the 
turbulent pressure (\nocite{xetal96}Xie \etal 1996).
Advances have also been made in modeling the 
dynamical evolution of \UCHII\ regions in the
presence of high pressure molecular environment. 
Through simulations, \nocite{gf04}Garcia-Segura \& Franco (2004)
show that pressure-confined \UCHII\ regions with ionizing stars
moving with respect to the molecular core can be long-lived 
and can produce the observed morphologies of the \HII\ regions.

If \UCHII\ regions are pressure confined due to
high density gas in their vicinity,
then other observable effects will occur.
In particular, far-ultraviolet (FUV) photons 
(6.0 -- 13.6 eV) from the OB star would produce photo-dissociation regions 
(PDRs) in the neutral material close to the \UCHII\
region. Near the interface between the region where hydrogen is
ionized and the PDR, gas phase carbon will be ionized by FUV photons 
in the energy range 11.3 to 13.6 eV (see review by 
\nocite{ht97}Hollenbach \& Tielens 1997). The  
electron density in this layer is relatively high ($>$ 10$^3$ \cmthree) 
and gas temperatures can be in the range 300 -- 1000 K
(e.g. \nocite{nwt94} Natta, Walmsley \& Tielens 1994). These conditions 
are ideally suited for producing observable radio recombination lines (RLs)
of carbon. 

Physical properties of dense material near  \UCHII\ regions were 
earlier estimated from observations of high density molecular tracers. 
However, high line optical depth in these regions complicates the 
determination of the physical properties. We propose that 
multi-wavelength carbon RL observations 
from PDRs associated with \UCHII\ regions can be used to 
estimate the properties of the dense molecular material. 
Unlike molecular traces, RLs do not suffer from the 
limitation of high opacity and thus provide another probe to 
test whether \UCHII\ regions are pressure confined.

In this paper, we present multi-wavelength (0.7, 2, 3.6 \& 6 cm) 
VLA observations of carbon RLs
from the PDR associated with G35.20$-$1.74. The \UCHII\ region 
is in the molecular cloud complex W48 at $l = $ 35\ddeg.2,
$b =$ $-$1\ddeg.74 at a distance of 3.2 kpc 
(\nocite{wc89b}Wood \& Churchwell 1989b). At 0.7 cm, we have 
detected the C53$\alpha$ and X53$\alpha$ (see \S\ref{sec:cline}) transitions.
This data forms the first
successful imaging of carbon RL at 0.7 cm with the VLA at an angular
resolution of about 2\arcsec.  The details of the observations 
are given in \S\ref{sec:obs}. In \S\ref{sec:cont} and 
\S\ref{sec:cline}, we discuss the modeling of the multi-wavelength data
to determine the physical properties of the PDR and the \UCHII\ region.
In \S\ref{sec:pres}, we estimate the gas pressure
in the \UCHII\ region and the PDR and investigate whether 
the \HII\ region is pressure confined.

\section{Observation and data reduction}
\label{sec:obs}

We made spectroscopic observations of W48 using the VLA in 
D-configuration at 0.7, 2, 3.6 and 6 cm in dual polarization mode. 
%The coordinates of the field center were 
%RA(2000) = 19$^h$01$^m$47$^s$.06 \& 
%DEC(2000) = +01\ddeg12\arcmin59\arcsec.6 for the 3.6 and 6 cm
%observations and RA(2000) = 19$^h$01$^m$46$^s$.4 \& 
%DEC(2000) = +01\ddeg13\arcmin23\arcsec.9 for the 2 cm observation.
The 0.7 cm observations were made during August 2004 and data at
other wavelengths were obtained during October 2001.
The transitions observed are the C110$\alpha$ (4876.5886 MHz)
and C111$\alpha$ (4746.5497 MHz) at 6 cm, the C92$\alpha$ 
(8313.5279 MHz) at 3.6 cm, the C76$\alpha$ (14697.3141 MHz) at 2 cm,
and the C53$\alpha$ (42973.3984 MHz) at 0.7 cm. 
The bandwidth at 2 cm wavelength was 3.13 MHz, 
which corresponds to a total velocity coverage of
64 \kms. Thus in addition to the carbon line, roughly 75 \% of 
the He76$\alpha$ line profile falls within the observed bandwidth. 
Table~\ref{tab:obs} summarizes the parameters of the observations.

Data analysis was carried out using the Astronomical Image Processing
Software (AIPS). The default channel zero data was used 
for continuum calibration.  After satisfactory editing and 
calibration, the flag and calibration 
tables were transferred to the spectral data. The system
band shapes were obtained using the bandpass calibrator data with
the AIPS task BPASS. The estimated band shapes were used for 
bandpass calibration. Line free channels from the bandpass
calibrated data were used for estimating
the continuum emission which was then subtracted from the
spectral line data.
We used the task UVLSF for this purpose. The continuum uv data
created by UVLSF was used as the input to IMAGR to make the 
continuum images. The spectral cubes were also made with the AIPS
task IMAGR. The data at 0.7 cm were analyzed by taking into account
of the weighting provided by the on-line system for each visibility
measurement. GIPSY (Groningen Image Processing System) 
and AIPS++ software packages were used to further process 
(e.g. Gaussian fit) the spectral line data.   
 
\section{Continuum Emission}
\label{sec:cont}

\begin{figure}
\plotone{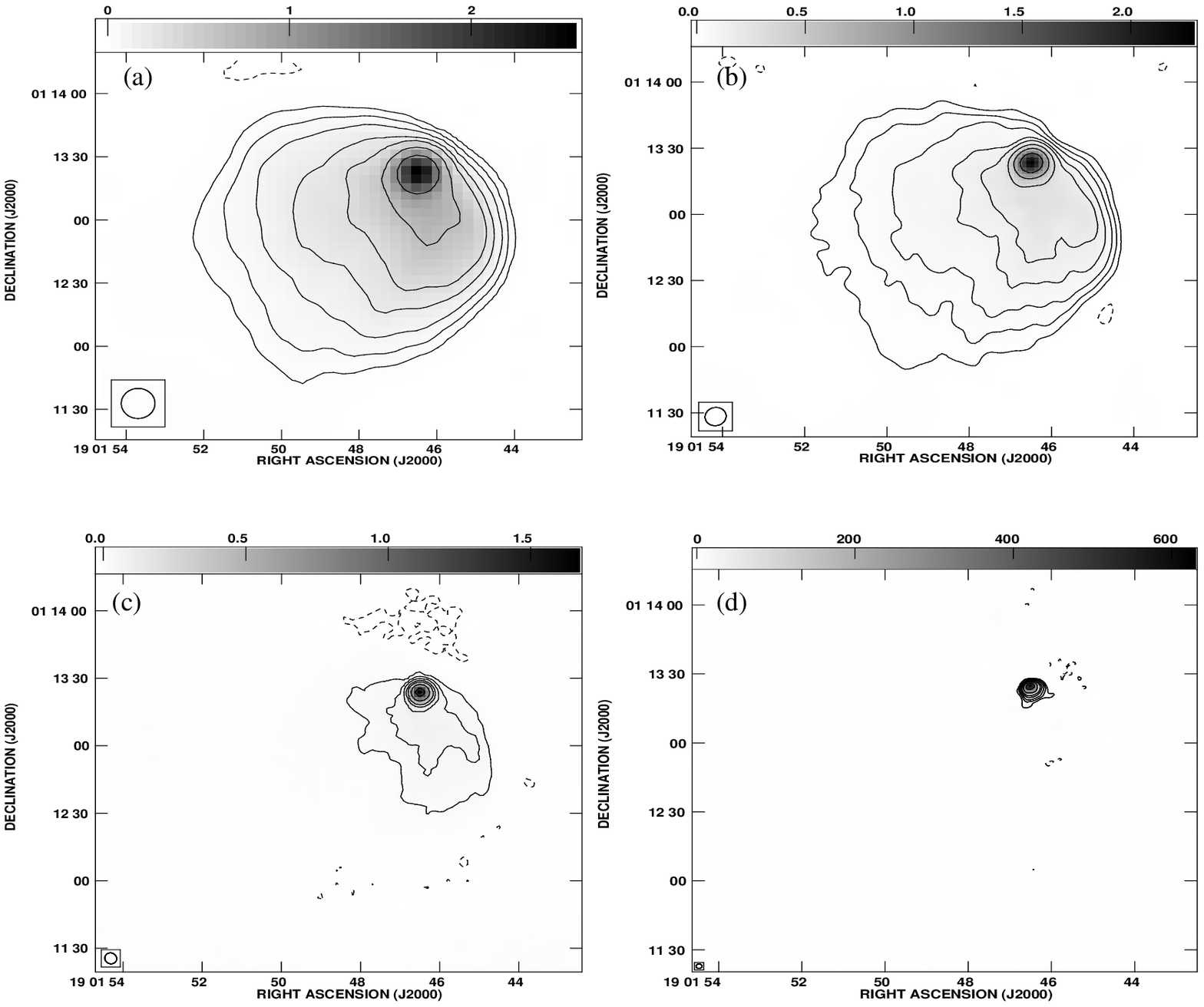}
\caption{
Continuum images of W48A (extended source) and the \UCHII\ region
G35.20$-$1.74 (compact source) made with the VLA. 
(a) 6 cm image with an angular resolution of 14\arcsec.3 $\times$  13\arcsec.0 
($-$0$^o$.8). The contour levels are 
($-$1, 2, 5, 10, 20, 32, 64, 128, 256) $\times$ 10 mJy/beam and
the grey scale ranges from $-$0.01 to 2.55 Jy/beam. 
The RMS noise in the image is 2.0 mJy/beam.
(b) 3.6 cm image with an angular resolution of 8\arcsec.4 $\times$  7\arcsec.6
($-$17$^o$.2). The contour levels are 
($-$1, 2, 5, 10, 20, 32, 64, 128, 256) $\times$ 5 mJy/beam and
the grey scale ranges from $-$0.01 to 2.29 Jy/beam. 
The RMS noise in the image is 1.4 mJy/beam.
(c) 2 cm image with an angular resolution of 4\arcsec.9 $\times$  4\arcsec.5
(17$^o$.7). The contour levels are 
($-$1, 2, 5, 10, 20, 32, 64, 128, 256) $\times$ 10 mJy/beam and
the grey scale ranges from $-$0.01 to 1.67 Jy/beam.
The RMS noise in the image is 2.9 mJy/beam.
(d) 0.7 cm image with an angular resolution of 2\arcsec.1 $\times$  2\arcsec.0
($-$89$^o$.9). The contour levels are 
($-$1, 2, 5, 10, 20, 32, 64, 128, 256) $\times$ 3 mJy/beam and
the grey scale ranges from $-$4 to 630 mJy/beam.
The RMS noise in the image is 0.7 mJy/beam.
\label{fig:1} } 
\end{figure}

%The molecular cloud complex W48 consists of several \HII regions
%in different stages of evolution. 
Fig.~\ref{fig:1} shows continuum images at 0.7, 2, 3.6 and 6 cm with 
angular resolutions ranging from $\sim$ 2\arcsec\ to 14\arcsec.
%of 14\arcsec.4 $\times$ 13\arcsec.0,
%8\arcsec.7 $\times$ 8\arcsec.1 and 5\arcsec.0 $\times$ 4\arcsec.6.
These continuum images show a compact source of 
angular size $\sim$ 5\arcsec\ $\times$ 5\arcsec\ along
with an extended source (angular size $\sim$ 1\arcmin.5 $\times$
1\arcmin.5), located south and east of the compact component 
(see Fig.~\ref{fig:1}). This extended source is designated as W48A in the 
literature (\nocite{opbgt94}Onello \etal\ 1994). It
has been suggested that the extended emission is directly associated with the 
compact component (\nocite{kf02}Kurtz \& Franco 2002). 
%This suggestion was made based on their
%continuum observations at 3.6 cm of the W48 star forming region. 
%However, we find that the central velocities of the 
%helium RLs (see \S\ref{sec:heline}) from the two sources differ 
%by 5.4$\pm$0.7 \kms. A velocity difference of 3.2$\pm$0.6 \kms\
%is also observed between the hydrogen (\nocite{opbgt94}Onello \etal 1994)
%and helium RLs from the two sources.  Therefore, in this paper, 
%we consider that the 
%two sources are not directly associated. 
%The extended source is 
%an evolved \HII\ region and is designated as W48A in the literature 
%(\nocite{opbgt94}Onello \etal\ 1994).

The continuum emission and the
partially ionized gas associated with W48A have been extensively 
studied earlier with a lower angular resolution of $\sim$ 60\arcsec\
%66\arcsec $\times$ 61\arcsec
(see \nocite{opbgt94}Onello \etal 1994). In our images, 
the emission from W48A extends to about 2\arcmin\ at the 
longest wavelength. The flux densities of the extended emission
measured are 5, 10 and 12 Jy respectively at 2, 3.6 and 6 cm
wavelengths. The measured flux density at 2 cm is about a
factor of two less than that estimated using the flux
densities at 3.6 and 6 cm wavelengths and considering a 
spectral index of $-$0.1 for the thermal emission.
% and assuming that the nebula is optically thin. 
This lower flux density is due to the missing short spacings
in interferometric observations. The extended emission is
not detected at 0.7 cm because of the same reason. 
The largest angular size that our 
observations are sensitive to are 5\arcmin, 
3\arcmin, 1.5\arcmin\ and 0.7\arcmin\ respectively at 6, 3.6, 2 and 0.7 cm. 
Therefore at 2 and 0.7 cm wavelengths the full extent and flux density of
the emission from W48A are not determined.  

%The compact source in Fig.~\ref{fig:1} is an \UCHII\ region 
%referred to as G35.2$-$1.74 
%in the literature (eg. \nocite{wc89}Wood and Churchwell 1989). 
%The continuum emission from G35.2$-$1.74
%resembles a cometary morphology (\nocite{wc89}Wood \& Churchwell 1989).
%The angular resolution of the observations (see Table~\ref{tab:obs}) 
%are not sufficient to resolve the emission from the \UCHII\ region. 
%Along with the course angular resolution, the extended emission 
%from W48A also makes it difficult to determine the 
%angular size and flux of the \UCHII\ region by a simple Gaussian
%fit to the continuum emission. Therefore we used the AIPS task
%SLICE to make profiles along RA and DEC. To these slices we fitted
%a Gaussian plus a 4$^{th}$ order polynomial, which models the extended
%emission. The flux determined from these fits 
%are given in Table.~\ref{tab:cs}.
%The uncertainties in the flux estimate given in the table
%are the RMS value of the residual
%of the model fits. The angular size of the source at 2cm, where the
%emission from W48A is weak, is obtained using the AIPS task IMFIT. 
%The flux and angular size are consistent with those estimated
%earlier. (\nocite{wc89}Wood \& Churchwell 1989, 
%\nocite{whp85}Woodward, Helfer \& Pipher 1985).

The compact source in Fig.~\ref{fig:1} is an \UCHII\ region 
referred to as G35.20$-$1.74 
in the literature (e.g. \nocite{wc89b}Wood and Churchwell 1989b). 
The continuum emission from G35.20$-$1.74 
shows some resemblance to a cometary morphology 
as inferred earlier by \nocite{wc89b}Wood \& Churchwell (1989b).
The flux densities of the compact source 
at the four observed wavelengths
are given in Table.~\ref{tab:cs}. 
%These values are obtained from slices made on the continuum
%images and then fitting a Gaussian plus 4$^{th}$ order polynomial, 
%which models the extended emission, to the slices.  
The angular size of the source is estimated
from the 2 and 0.7 cm images, which have the highest angular resolutions.
The AIPS task JMFIT was used to determine the angular size. 
The flux density and angular size are consistent with those estimated
earlier (\nocite{wc89b}Wood \& Churchwell 1989b, 
\nocite{whp85}Woodward, Helfer \& Pipher 1985).

\begin{figure}
\plotone{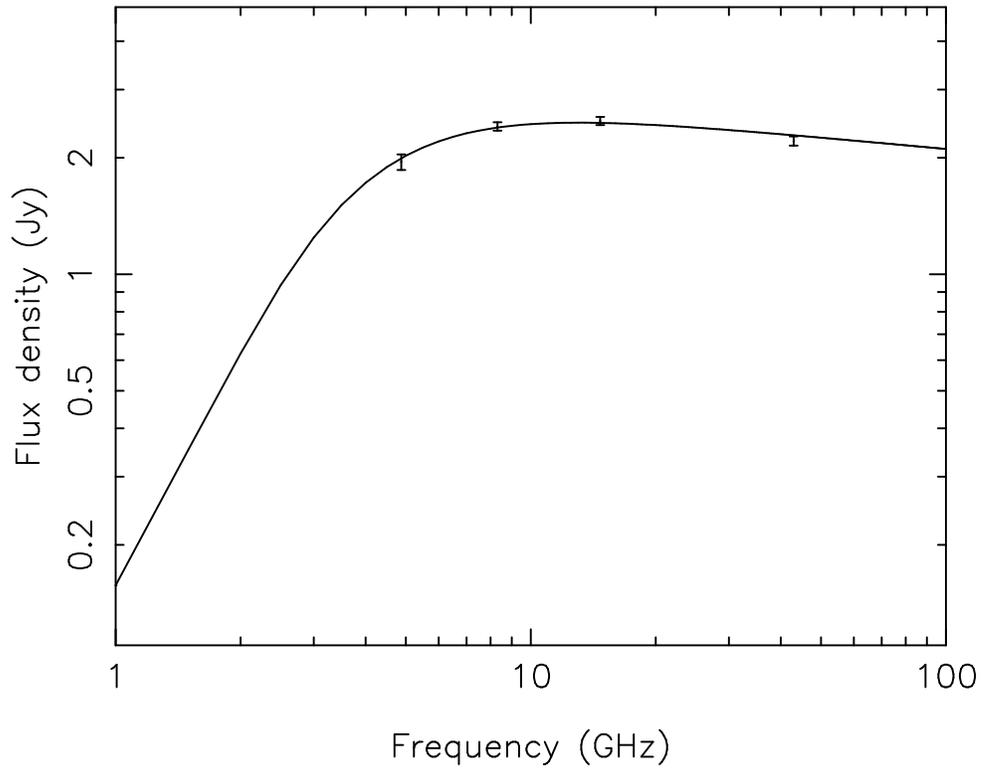}
\caption{
Flux density vs frequency of the \UCHII\ region G35.20$-$1.74. 
The solid curve shows the model flux density  as a function of 
frequency for $T_e = 9900 K$ and 
$EM = 6.4 \times 10^7$ pc cm$^{-6}$.  The measured flux density at
the four frequencies along with $\pm 3\sigma$ error bars are also marked.
\label{fig:2} } 
\end{figure}

The continuum data toward G35.20$-$1.74 are used to determine the 
physical properties of the \UCHII\ region. 
%Given the course angular resolution of the observations 
%it is not possible to model the cometary morphology accurately. 
We model the continuum emission by considering that the emission 
from the compact source originates from a spherical, homogeneous ionized region. 
The flux density due to thermal emission from such an ionized region
is estimated as described by \nocite{mh67}Mezger \& Henderson (1967).
%and compared it with the observed values.
The angular size of the spherical region estimated from
the observed size of the \UCHII\ region is 6\arcsec.3. 
%(= 1.724 $\times$ 3.65; see Table~\ref{tab:cs}). 
Table~\ref{tab:pp} gives the parameters
of the ionized gas obtained from continuum modeling.
A plot of the flux density from the model as a function of frequency
is shown in Fig.~\ref{fig:2}. 
At a distance of 3.2 kpc (\nocite{wc89b}Wood \& Churchwell 1989b), 
the diameter of the spherical region of ionized gas is 0.1 pc. 
The derived physical properties and the observed flux densities 
are then used to estimate the
excitation parameter, rate of Lyman continuum photons required to
maintain ionization equilibrium and the spectral type of the embedded
star (\nocite{p73}Panagia 1973). These values are also included in  Table~\ref{tab:pp}.

\section{Recombination Line Emission toward G35.20$-$1.74}
\label{sec:cline}

\subsection{Carbon Recombination Lines}
\label{sec:cline1}

\begin{figure}
\plotone{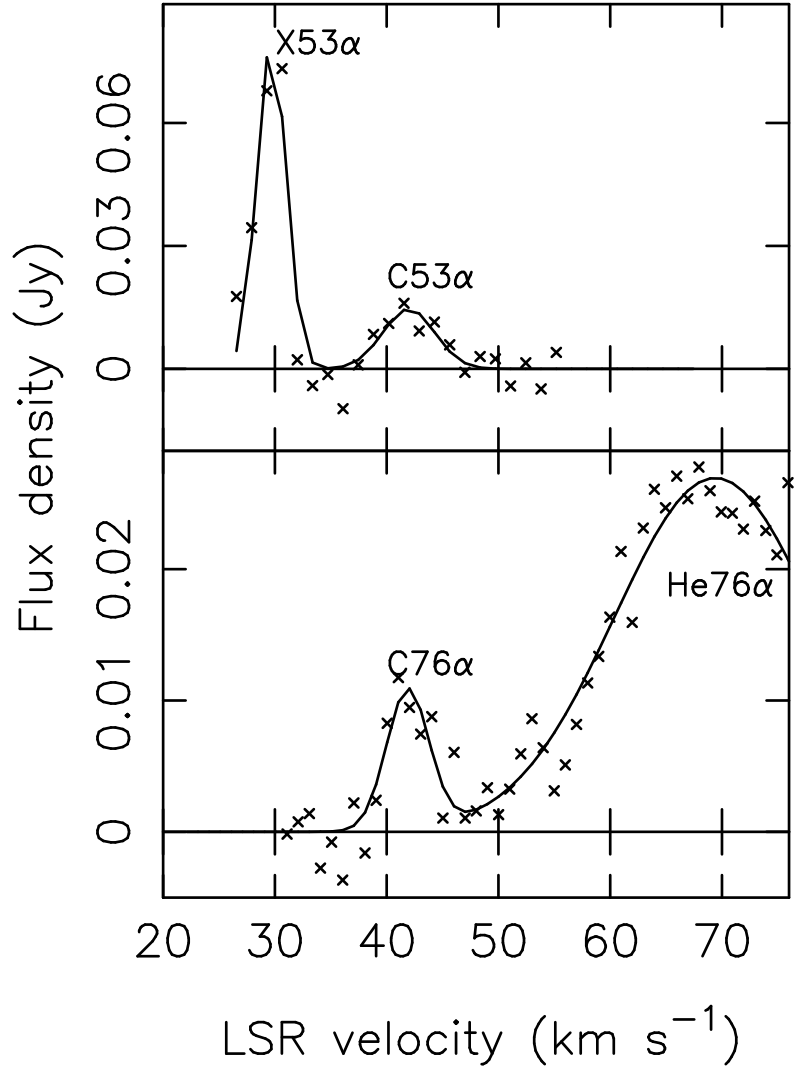}
\caption{
Spectra of the 53$\alpha$ and 76$\alpha$ recombination line transitions observed
near 2 (bottom) and 0.7 (top) cm toward G35.20$-$1.74. The spectra are obtained
by averaging over a 6\arcsec.3 $\times$ 6\arcsec.3 region near  
the continuum peak (RA = 19$^h$01$^m$46$^s$.4, 
DEC = +01$^{o}$13$^{'}$24$^{''}$, J2000) in the 2 and 0.7 cm images. 
The line features near $+$42 \kms\ are the carbon lines. 
The He76$\alpha$ with an LSR velocity of $+$42.0$\pm$0.5 \kms\
is the second feature in the spectrum shown on the bottom panel.
Only part of the helium line profile is detected due to limited 
bandwidth of the observation. The line feature near $+$30 \kms\ 
in the top spectrum is presumably a heavy ion (ion massive that C$^{+}$) 
recombination line. The LSR velocity is with respect to the carbon lines.
The result of a two component Gaussian fit to the line features 
is shown by the solid line. 
%The residual after removing the 
%Gaussian components are shown by the points. 
%The y-axis is the 
%flux density measured with a beam equal to the angular
%resolution of 5\arcsec.0 $\times$ 4\arcsec.6. 
\label{fig:3} }
\end{figure}

We detected the C53$\alpha$ and C76$\alpha$ recombination lines in 
the direction of G35.20$-$1.74. No carbon RLs were detected toward the 
extended \HII\ region W48A. The non-detections toward other directions 
indicate that the carbon line originates from 
the PDR associated with the \UCHII\ region.
The spectra at 2 and 0.7 cm wavelengths, averaged 
over a 6\arcsec.3 $\times$ 6\arcsec.3  
region near the continuum peak toward G35.20$-$1.74, are 
%(RA = 19$^h$01$^m$46$^s$.4, 
%DEC = +01$^{o}$13$^{'}$23$^{''}$.9, J2000) 
shown in Fig.~\ref{fig:3}
and the parameters obtained from the Gaussian fit to the line features 
in the figure are given in Table.~\ref{tab:line}. 
The upper limits on the carbon line emission at 3.6 and 6 cm are  
included in Table.~\ref{tab:line}. The data at 3.6 cm are affected
by a systematic baseline ripple that was not removed by a careful 
bandpass calibration. 
%The origin of this ripple could not 
%be traced, although we have spent a lot of time in carefully 
%editing the data. 
This ripple is present in the
spectra toward regions with bright continuum emission. 
The upper limit we obtained toward the UCHII region is 
from the spectrum with the baseline ripple and
therefore its value is higher (5.2 mJy/beam) compared to that estimated 
from an off source position (= 1.1 mJy/beam). 
%The upper limit of 5.2 mJy at 3.6 cm is obtained from the spectrum 
%toward the \UCHII\ region, which also had the baseline ripple. Therefore 
%the estimated upper limit is higher than the RMS (= 1.1 mJy) 
%obtained from offsource position. 

%The Stokes I spectrum at 3.6 cm toward the same position showed a 
%line like feature. However,
%the reality of this feature could not be confirmed by independently
%detecting the line in the two polarizations even after 
%extensive examination and editing
%of data. Therefore we consider the line like feature as spurious.
%The upper limit on the carbon line emission at 3.6 and 6 cm are  
%included in Table.~\ref{tab:line}. Note that the RMS at 3.6 cm
%is estimated from the spectrum containing the spurious feature.

\subsection{Models for the Carbon line emission toward G35.20$-$1.74}
\label{sec:cmodel}

We consider homogeneous `slabs' of PDR material  
placed in front and back of the \UCHII\ region and solved the
radiative transfer equation for non-LTE cases to obtain
the recombination line flux density. 
The non-LTE departure coefficients $b_n$ and $\beta_n$ 
are calculated using the program
originally developed by \nocite{bs77}Brocklehurst \&  Salem (1977)
and later modified by \nocite{ww82}Walmsley \& Watson (1982) and 
\nocite{pae94}Payne, Anantharamaiah \& Erickson (1994).
The two relevant input parameters for the program are: 
(1) abundance of gas phase carbon and (2) background radiation field.
The abundance of carbon is taken as 0.75 of the standard abundance 
(3.9 $\times$ 10$^{-4}$; \nocite{m74}Morton 1974), which 
implies a depletion factor of 25 \% (\nocite{nwt94}Natta \etal 1994).
The departure coefficients depend on the
background radiation field. We have selected a thermal background 
due to an \UCHII\ region with temperature and emission measure
determined from the continuum observations (see 
\S\ref{sec:cont}).
The departure coefficients are computed for a set of electron 
temperatures in the range 100 to 1000 K and densities
between 1000 and 6000 \cmthree. 
The dielectronic like recombination process that 
modifies the level population (\nocite{ww82}Walmsley \& Watson 1982) 
is also included 
in the calculation. The coefficients are computed by considering 
a 10000 level atom with the boundary condition 
$b_n \rightarrow 1$ at higher quantum states (see 
\nocite{pae94}Payne \etal 1994
for further details).  

The line intensity is a function of PDR gas temperature, $T_{PDR}$, 
electron density, $n_e^{PDR}$, PDR thickness along the line of sight, 
$l$, and the background radiation field. For the homogeneous PDR,
the line brightness temperature, $T_{LB}$, due to the slab in the near side
of the \UCHII\ region is given by (\nocite{s75}Shaver 1975)
\begin{equation}
T_{LB}  =  T_{bg,\nu} + T_{in,\nu}\ ,
\label{eq:line1}
\end{equation}
where
\begin{eqnarray}
T_{bg,\nu} & = & T_{0bg,\nu} e^{-\tau_{C\nu}}(e^{-\tau_{L\nu}} - 1) \ ,\nonumber \\
T_{in,\nu} & = & T_{PDR}\left( \frac{b_m \tau_{L\nu}^* + \tau_{C\nu}}{\tau_{L\nu} + \tau_{C\nu}}
          (1 - e^{-(\tau_{L\nu} + \tau_{C\nu})}) - (1 - e^{-\tau_{C\nu}})\right). 
\label{eq:line}
\end{eqnarray}
$T_{bg,\nu}$ is the contribution to the line temperature due to the
background radiation field and $T_{in,\nu}$ is the intrinsic line emission from
the slab. In Eq.~\ref{eq:line},  $T_{0bg,\nu}$ is the background 
radiation temperature, which, as
discussed above, is the continuum emission from the \UCHII\ region.  
$\tau_{C\nu}$ is the continuum optical depth of the PDR. The non-LTE line 
optical depth of the spectral transition from energy state $m$ to $n$, 
$\tau_{L\nu}$, is $\tau_{L\nu}  =  b_n \beta_n \tau_{L\nu}^*$
where $\tau_{L\nu}^*$ is the LTE line optical depth,  
$b_n$ and $\beta_n$ are the departure coefficients of state $n$. For the
PDR, $\tau_{L\nu}^*\; \propto\;  n_e^{PDR} n_{C^+} l$, where $n_{C^+}$ is the carbon
ion volume number density in the PDR. For the present calculations we assumed 
that $n_e^{PDR} = n_{C^+} = n_e$ so $\tau_{L\nu}^*\; \propto\;  n_e^2 l$.
The line temperature from the PDR on the far side (see below) is obtained 
from Eq.~\ref{eq:line1} by setting $T_{bg,\nu} = 0$. 
The line brightness temperature is
finally converted to flux density using an angular size of 
about 3\arcsec.7  (see Table~\ref{tab:cs}).

%For a given electron temperature, we varied the emission measure to obtain the line 
%flux density at 0.7 and 2 cm. 

%The source angular size used to estimate the
%line flux density is same as the 2 cm synthesized beam.
%The synthesize beam is used because the modeling
%is done using the flux density toward the continuum peak at 2 cm.
%The line temperature 
%is then converted to flux using the equation
%\be
%S_L = 3.1 \times 10^{-14} T_L \Omega_s \nu^2
%\ee
%where $S_L$ is the line flux in Jy, $T_L$ is the line temperature in K,
%$\Omega_s$ is the source size in strad and $\nu$ is the
%frequency in Hz. 

\begin{figure}
\plotone{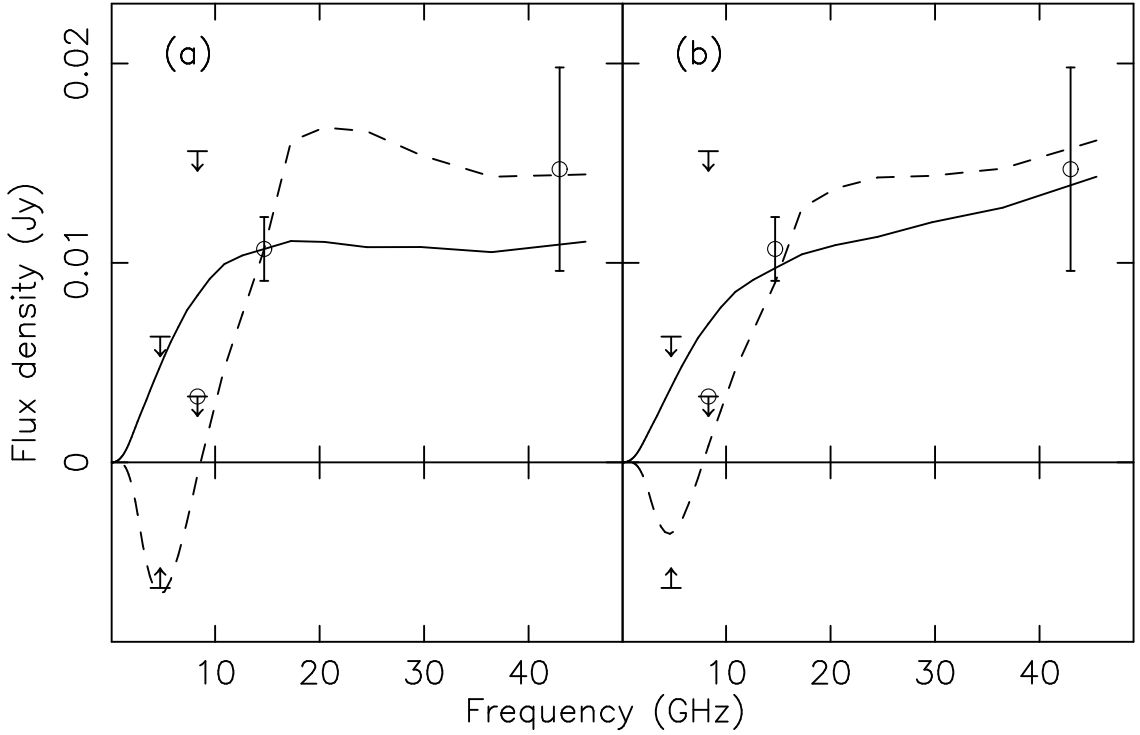}
\caption{
Carbon line flux density as a function of frequency for class (a) and (b)
models (see \S\ref{sec:cmodel}) are shown on Figs. (a) and (b) respectively.
The solid and dashed curve corresponds to gas 
temperatures of 500 and 200 K respectively. The electron density used to
obtain all curves are 2500 \cmthree. The upper limits at 6 and 3.6 cm
wavelengths and the line flux densities of C53$\alpha$ and C76$\alpha$ 
transitions with $\pm$ 1$\sigma$ error are also marked. At frequencies
below $\sim$ 10 GHz, the 200 K model predicts the carbon RLs in absorption.
The parameters of such models are constrained by using the 6 cm spectral
uncertainty (3$\sigma$) as a lower limit for the model flux densities 
at this wavelength, which is also shown in the figure.  At 3.6 cm the
upper limit obtained from off source spectra (3.3 mJy) is shown
with a circle. Note that the 200 K model flux densities are consistent with
this upper limit. 
\label{fig:4} }
\end{figure}

We considered three classes of models -- (a) models with line emission 
due to a PDR slab in the front of the \UCHII\ region; (b)   
models with line emission having contribution from  
PDR slabs in the front and back of the \UCHII\
region and (c) model with line emission due to a PDR slab in
the far side of the \UCHII\ region. The line emission for class (a) models
are obtained using Eqs.~\ref{eq:line1} and \ref{eq:line}. For
class (b) models, the intrinsic contribution (ie $T_{in,\nu}$ in
Eq.~\ref{eq:line1}) from the PDR slab on the far side is
added to Eq.~\ref{eq:line1}. Note that in this case we   
essentially assume that the relative motion between the
PDR slabs in the front and back of the \UCHII\ region is smaller
than the width of the observed carbon line (see \S\ref{sec:relv}). 
For each class of models, a set of parameter values is determined such 
that the observed line flux densities at 0.7 and 2 cm (also consistent 
with the observed upper limits at 3.6 and 6 cm) are reproduced within 
$\pm$ 1$\sigma$ error. We found that
class (c) models cannot reproduce the observed line flux densities and hence
will not be discussed further. Table~\ref{tab:pdrp}
gives a subset of model parameters that are consistent with our RL data
at the four wavelengths. 
The neutral densities listed in Table~\ref{tab:pdrp} are obtained with a 
gas phase carbon abundance of 3 $\times$ 10$^{-4}$ used for modeling. 
Fig.~\ref{fig:4} shows the model 
carbon line flux density as a function of frequency
along with the observed values and limits for class (a) and (b) models 
with typical temperatures 200 and 500 K and electron density 2500 \cmthree.

The modeling shows that: (1) the electron density and PDR thickness
are well constrained by our RL data with the density in  
the range 1500 -- 6000 \cmthree and the PDR thickness 
a few times 10$^{-4}$ pc; (2) models with temperatures $<$ 150 K  
are ruled out by our RL data. We have explored models
with PDR temperatures up to 1000 K (\nocite{nwt94}Natta \etal 1994).
However, the temperature could not be well constrained due to
the uncertainty in the upper limit obtained from the 3.6 cm spectrum
(see \S\ref{sec:cline1}). Note that models with temperature $\le$ 250 K are
consistent with the 3$\sigma$ upper limit obtained from the off source
region of 3.3 mJy at 3.6 cm (see Fig.~\ref{fig:4}). 

In a recent single dish survey of carbon lines near 
8.5 GHz, \nocite{retal05}Roshi \etal\ (2005) 
detected RLs toward a large number of \UCHII\ regions. These lines
along with upper limits at other frequencies were used to model the
properties of the line forming region. The ranges of density and
size obtained for the PDR toward G35.2$-$1.74 are comparable with
those determined by \nocite{retal05}Roshi \etal\ (2005) toward 
other \UCHII\ regions.

An important inference from modeling is that the line emission 
at frequencies near $\sim$ 14 GHz is dominated by stimulated 
emission due to the background continuum emission arising 
from the \UCHII\ region. The intrinsic line flux density from 
the slab near 14 GHz is $<$ 20 \% of the observed RL flux density. 
Thus for these frequencies Eq.~\ref{eq:line1}
can be approximated as $T_{LB} \sim T_{bg,\nu}$. 
At frequencies \gsim 14 GHz, the intrinsic line emission
from the slab (ie $T_{in,\nu}$ in Eq.~\ref{eq:line1}) contributes
significantly. For example, at 0.7 cm the intrinsic line flux density
from the PDR slab is about 6 mJy for models with electron
density 3000 \cmthree, which is almost equal to the line flux density
due to the background term (ie $T_{bg, \nu}$) in Eq.~\ref{eq:line1}. 

%The dominance of stimulated emission
%implies that at 2 cm the line emission is detected only
%from the PDR in front of the \UCHII\ region. In addition to
%the effect of stimulated emission, the line emission from 
%the PDR behind 
%the \UCHII\ region will be attenuated due to the higher
%opacity of the ionized gas at larger wavelengths. The 
%continuum optical depth at 0.7, 2, 3.6 and 6 cm wavelengths 
%are 0.01, 0.08, 0.3 and 0.8 respectively. 
%The expected carbon line flux density at 2 cm from the PDR behind the 
%\UCHII\ region is $<$ 2 mJy. At 0.7 cm, the intrinsic line emission
%from the PDR slabs contribute as much as   
%For models with electron density 
%$>$ 3000 \cmthree, line flux density at 0.7 cm 
%from the PDR at the far side is about 6 mJy. 
%Thus the sensitivity
%of our observations is not sufficient to detect the line
%emission from the PDR behind the \UCHII\ region at both wavelengths. 

\subsection{Helium and Other Recombination Lines}
\label{sec:heline}

In addition to the C76$\alpha$ transition, 
He76$\alpha$ RLs in the direction of G35.20$-$1.74 
and W48A were detected.  The helium line parameters are 
given in Table~\ref{tab:line}. The LSR velocity of the
He76$\alpha$ line ($+$42.0$\pm$0.5 \kms) differs by $-$5.9 \kms\
from the earlier measured value for the H76$\alpha$ line
(\nocite{wc89b}Wood \& Churchwell 1989b). The parameters
obtained for the H76$\alpha$ RL are $V_{LSR}$ = $+$47.9$\pm$1.2 \kms,
$\Delta V$ = 31.8$\pm$1.89 \kms\ and $S_L$ = 340$\pm$26 mJy. The
angular resolution of our 2 cm observation is similar to that of
the earlier observations ($\sim$ 4\arcsec ; \nocite{wc89b}Wood 
\& Churchwell 1989b). Differences in the sizes of the regions 
where helium and hydrogen are ionized could be the cause of
the LSR velocity difference between the two RLs. The 
limited velocity coverage of the 2 cm data (see \S\ref{sec:obs}) 
could also be the cause of the LSR velocity difference.  
Further confirmation is needed however. 

The line emission from
W48A is averaged over an area of 36\arcsec\ (in RA) $\times$ 17\arcsec\ (in Dec) 
centered at  RA(2000) 19$^h$01$^m$46$^s$.4 and 
DEC(2000) 01$^{o}$13$^{'}$02$^{''}$. 
%RA(2000) 19$^h$01$^m$47$^s$.6 and 
%DEC(2000) 01$^{o}$12$^{'}$55$^{''}$.4 to RA(2000) 19$^h$01$^m$45$^s$.2 and
%DEC(2000) 01$^{o}$13$^{'}$12$^{''}$.3.  
This region does not include the \UCHII\ region and 
thus is an estimate of the helium
line emission from W48A. The line parameters given
in Table~\ref{tab:line} are from the average spectrum.
The line flux density from W48A is only about 6 \% of 
that toward the \UCHII\ region. 
%The central velocity of the helium line from
%W48A is $+$5.4 \kms\ higher than that of the He76$\alpha$ RL
%$\pm$0.7 
%toward the \UCHII\ region. 
%The LSR velocities of the helium lines toward 
%W48A and G35.20$-$1.74 are $+$47.4 and $+$42.0 \kms\ respectively.
The LSR velocity of the helium line from W48A ($+$47.4$\pm$0.5 \kms) 
is similar to that of the H76$\alpha$ line detected from G35.20$-$1.74 by 
\nocite{wc89b}Wood \& Churchwell (1989b). 

At 0.7 cm wavelength, a line feature at LSR velocity about 
$+$43.4 \kms\ is detected. The LSR velocity is obtained by
assuming that the mass of the ion to be infinity.
\nocite{v87}Valee (1987) have detected the S125$\alpha$ recombination line 
toward W48A. However, the LSR velocity of the sulfur line ($+$43.1 \kms; 
\nocite{v87}Valee 1987) is similar to that of carbon lines detected
in our observations. Therefore the second line feature detected at 0.7 cm 
wavelength may not be a sulfur line. The possibility of this line being
a Doppler shifted carbon RL cannot be ruled out. Further investigation 
is needed to identify this line feature. 

\section{Is G35.20$-$1.74 pressure confined ? }
\label{sec:pres}

As discussed in \S\ref{sec:intro} the lifetime of \UCHII\ regions
can be extended to a few times 10$^5$ years if they are pressure confined.
In this section, the derived physical properties 
of gas inside and outside G35.20$-$1.74 are used to investigate 
whether the \UCHII\ region is in fact pressure confined.

\subsection{Estimation of gas pressure} 

The total gas pressure inside the \UCHII\ region is the sum of thermal 
pressure and turbulent pressure. 
%Because there are no direct 
%measurements of the magnetic field strength in \UCHII s or PDRs 
%we do not include any contribution to the pressure from magnetic fields.
%The magnetic fields inside
%\HII\ regions are typically $\sim$ 10 -- 20 $\mu$G 
%(e.g. \nocite{hc80}Heiles \& Chu 1980) and hence do not contribute 
%significantly to the total pressure. 
No measurement of the magnetic field in G35.20$-$1.74 or the PDR 
associated with the \HII\ region exists and hence
we do not include its contribution in the calculation of total pressure. 
The total gas pressure is given by
\begin{equation}
P_{uchii} = 2 k n_e T_e + n_e \mu m_H v_{H-turb}^2 \mbox{~~dyne cm$^{-2}$},
\label{eq:pres}
\end{equation}
where $k$ is the Boltzmann constant, $n_e$ is the electron density in \cmthree,
$T_e$ is the electron temperature in K, $m_H$ is the hydrogen mass in gm 
and $v_{H-turb}$ is the turbulent velocity
inside the \UCHII\ region in units of cm s$^{-1}$. 
The line profile due to turbulence is considered to be
Gaussian with $\sigma = v_{H-turb}$, which is
estimated from the observed FWHM of 
H76$\alpha$ transition (\nocite{wc89b}Wood \& Churchwell 1989b) 
and the calculated line width due to thermal motion.  
The effective mass in amu of H + He gas with He fraction 
in number of atoms taken as 10 \% of that of H is 1.4. 
From the flux densities of H76$\alpha$ and He76$\alpha$ RLs we
infer that about 8 \% of the helium is ionized. 
Thus the parameter $\mu$ in the turbulent
pressure term in Eq. \ref{eq:pres} is $\sim$ 1.3 
since it is expressed in terms of $n_e$.  
The values for electron density and 
temperature, estimated from modeling the continuum emission from
the \UCHII\ region, are used to estimate gas pressure. 
%\be
%v_{He-turb}  =  \sqrt{v_{width}^2 - v_{thermal}^2}, \\
%\ee
%\be
%v_{thermal}  =  0.21 \sqrt{\frac{T_e}{M_{amu}}}~\mbox{km s$^{-1}$},
%\label{eq:vtherm}
%\ee
%where $v_{width}$ is the observed line width of He76$\alpha$ transition, 
%$v_{thermal}$ is the line width
%due to thermal motion, which in \kms\ is given by 
%Equation~\ref{eq:vtherm}. $M_{amu}$ 
%is the atomic mass in amu, which for He atom is 4.0. The estimated 
%turbulence width is 21 \kms. 
The total pressure inside the \UCHII\ region is 
1.3 $\times$ $10^{-7}$ \dycm, with a contribution 
of $7.0 \times 10^{-8}$ \dycm from thermal process and 
$5.7 \times 10^{-8}$ \dycm from turbulence.  The 
turbulent pressure is about 85 \% of the thermal pressure. 

The total gas pressure in the PDR is given by
\be
P_{PDR} = k n_H T_{PDR} + n_H \mu m_H v_{C-turb}^2 \mbox{~~\dycm},
\ee
where $n_H$ is the number density of hydrogen atoms in \cmthree, 
$T_{PDR}$ is the PDR gas temperature in K, $v_{C-turb}$ is the 
turbulent velocity in cm s$^{-1}$.  The line profile due to 
turbulence is considered to be Gaussian with $\sigma = v_{C-turb}$, 
which is estimated from the observed FWHM of C76$\alpha$ transition.
The effective mass $\mu$ in amu is taken as 1.4.
To estimate PDR pressure we consider that in the region where carbon is 
ionized, the neutral material is predominantly in atomic hydrogen form
and the gas temperature is equal to the electron temperature estimated
from carbon line modeling (see \nocite{ht97}Hollenbach \& Teilens 1997). 
%The turbulence velocity is estimated 
%from the carbon line width as described above for the 
%case of the helium line. 
%In the periphery of \UCHII\ regions the magnetic field 
%can be high (a few mG; e.g. \nocite{zrm00}Zheng, Reid \& Moran 2000) and 
%hence this contribution to the total pressure 
%cannot be neglected. However, no  measurement of the 
%magnetic field in the case of G35.20$-$1.74 exists and hence
%No measurement of the magnetic field in the PDR associated with
%G35.20$-$1.74 exists and hence
%we do not include its contribution in the calculation of PDR pressure. 
The estimated total gas pressure in the PDR is between 5.3 $\times$ 10$^{-7}$
and 4.3 $\times$ 10$^{-6}$  \dycm for the model parameters given in
Table~\ref{tab:pdrp}. For PDR temperatures $\le$ 500 K, the turbulent 
pressure is at least 10\% more than the thermal pressure.
This importance of turbulent pressure was noted earlier by 
\nocite{xetal96}Xie \etal (1996).

Comparing gas pressures in the \UCHII\ region and the PDR indicates
that the pressure in the PDR is at least four times larger than that 
in the \UCHII\ region. In \S\ref{sec:relv}, we investigate 
the relative motion between the different components 
(ionized gas, material in the PDR and molecular gas) 
along the line-of-sight to understand this pressure difference. 

%This relatively high pressure in the PDR is unexpected. 
%If such high pressure difference were present, relative motion 
%between the different 
%components (ionized gas, material in the PDR and molecular gas) 
%along the line-of-sight would be expected. 

\subsection{Kinematics}
\label{sec:relv}

\begin{figure}
\plotone{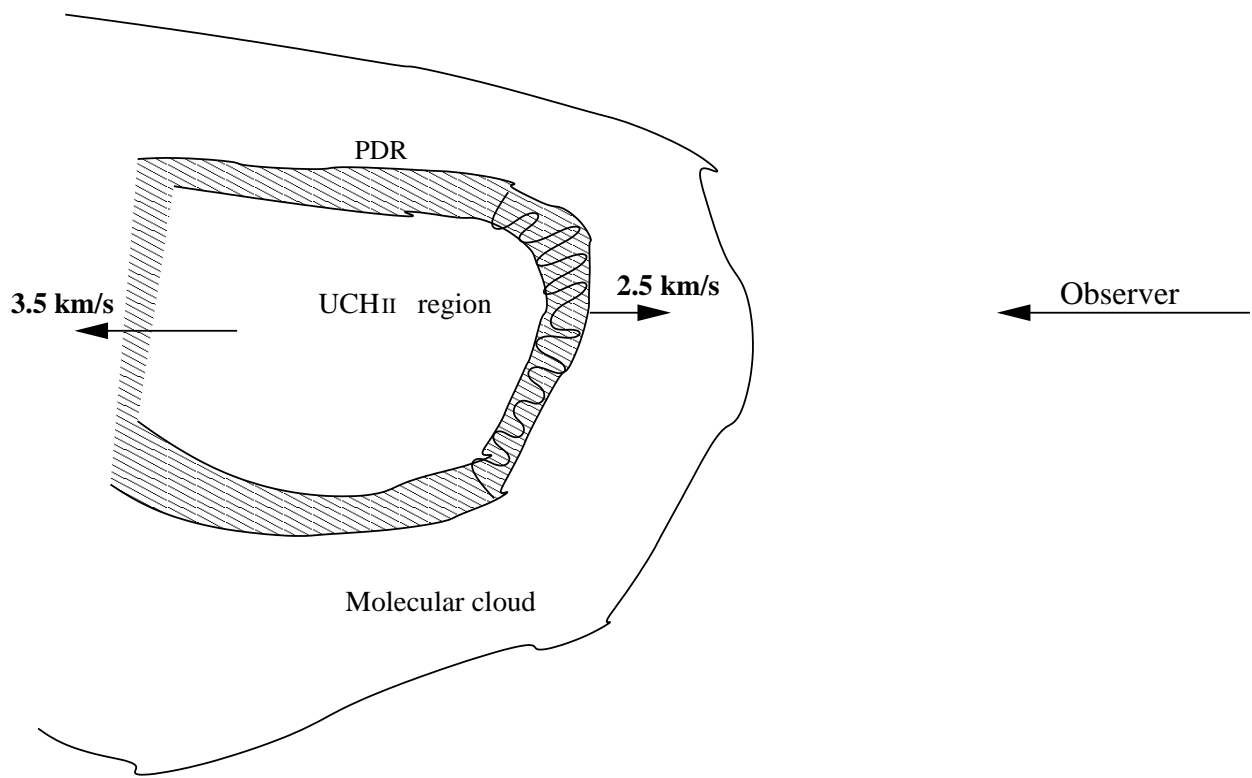}
\caption{
Schematic of the \UCHII\ region G35.20$-$1.74, the associated PDR 
and the molecular cloud.
The hydrogen RL at 14.6 GHz originates from the \UCHII\ region. The
central velocity of the H76$\alpha$ transition ($+$47.9$\pm$1.2 \kms) 
indicates the mean velocity of the ionized 
gas with respect to the LSR. The carbon RL observed at 14.6 GHz 
is detected from the region of the PDR marked by the 
curving line, since the line emission is dominated by stimulated emission. 
The central velocity of C76$\alpha$ line   
($+$41.9$\pm$0.4 \kms). The mean LSR velocity of the observed molecular
lines is $+$44.4$\pm$1.0 \kms. Thus relative to the molecular cloud the PDR is
moving into the cloud at $-$2.5$\pm$1.1 \kms\ and the ionized gas is moving at
$+$3.5$\pm$1.6 \kms. 
\label{fig:5} }
\end{figure}

The \UCHII\ region G35.20$-$1.74 is located in the molecular cloud
complex W48. The carbon line emission originates from PDR formed at
the interface between the molecular cloud and the \HII\ region. 
The PDR resides within the molecular cloud. The relative motion
along the line-of-sight between the three components (ionized gas, 
material in the PDR and molecular gas) can be determined by examining 
the central velocities of tracers of the different components
(see Fig.~\ref{fig:5}). The central velocities of high density 
tracers of molecular gas observed toward G35.20$-$1.74 and the angular
resolution of these observations are given in Table~\ref{tab:hdvlsr}.
The measured velocities with respect to the Local Standard of Rest (LSR) 
ranges from  $+$42.8 to $+$46.1 \kms. 
(The errors in the central velocities are not given in the literature
except for NH$_3$ (2,2), which is 0.1 \kms.) 
A variety of causes such as turbulence, shocks, complex excitation and 
chemistry in the vicinity of the \UCHII\ region contribute to the 
spread in the central velocity. Moreover, the different angular 
resolutions of the molecular line observations
may also contribute to the velocity difference. Molecular
line observations with higher angular resolution may help to precisely
determine the LSR velocity of the cloud associated with 
the \UCHII\ region. Here we take the mean value 
($+$44.4$\pm$1.0 \kms) of the central velocities listed in 
Table~\ref{tab:hdvlsr} as the LSR velocity of the molecular cloud.

%From the existing data we estimate that the relative velocity between 
%the ionized gas and the molecular cloud is \lsim 4.2 \kms. 

%In this section, we investigate whether a large pressure difference
%exists by examining the central velocities of the tracers of 
%different components along the line-of-sight toward G35.20$-$1.74
%(see Fig.~\ref{fig:5}).
%This study is used to investigate the existence of relative motion 
%between the components if large pressure difference do exist.

The H76$\alpha$ line (\nocite{wc89b}Wood \& Churchwell 1989b), 
which originates from the ionized gas 
in the \UCHII\ region (see \S\ref{sec:heline}), has 
a LSR velocity of $+$47.9$\pm$1.2 \kms.  Since the \UCHII\ region is
optically thin at 2 cm, this velocity provides the mean velocity of 
the ionized gas with respect to the LSR. 
As discussed in \S\ref{sec:cmodel}, the carbon RL flux density at 2 cm 
is dominated by stimulated emission and hence we preferentially
observe the emission from the PDR in front 
(ie toward the observer; see Fig.~\ref{fig:5}) 
of the \UCHII\ region. The carbon RL at 2 cm has a central velocity 
of $+$41.9$\pm$0.4 \kms. 

Translating the measured radial velocities to a reference frame
at rest with respect to the molecular cloud gives the  
picture of the PDR moving into the molecular cloud at 
$-$2.5$\pm$1.1 \kms\
and the ionized gas moving at $+$3.5$\pm$1.6 \kms\ relative to the
molecular cloud (see Fig.~\ref{fig:5}). 

%As discussed in Section~\ref{sec:heline}, the He76$\alpha$ line 
%originates from the ionized gas in the \UCHII\ region. 
%At 2 cm the \UCHII\ region is optically thin (see Section~\ref{sec:cont}). 
%Therefore the central velocity provides the mean velocity of 
%the ionized gas with respect to the Local Standard of Rest (LSR). 
%The LSR velocity of the 
%He76$\alpha$ line is $+$42.2$\pm$0.5 \kms (see Table~\ref{tab:line}). 
%The C76$\alpha$ line originates from
%the PDR adjacent to the \UCHII\ region. The 
%line emission at this frequency is dominated by stimulated emission. 
%Hence the C76$\alpha$ line originate from the the PDR in front of 
%the \UCHII\ region (see Section~\ref{sec:cmodel}). 
%The LSR velocity of C76$\alpha$ line is $+$42.3$\pm$0.5 \kms. 

\subsection{Discussion}
\label{sec:sum}

Modeling the multi-wavelength ($\lambda = $ 0.7, 2, 3.6 \& 6 cm) 
carbon recombination line and continuum data toward the \UCHII\ region
G35.20$-$1.74 has provided the physical properties of 
the ionized gas and the PDR associated with the \HII\ region. 
These physical properties were then used to determine 
the pressure in the \UCHII\ region and the PDR in order 
to investigate whether the \UCHII\ region
is pressure confined. Our calculations have shown that the 
pressure in the PDR is at least four times 
larger than that in the \UCHII\ region. Does this large
pressure in the PDR indicate that the \UCHII\ region
is pressure confined ? As shown in \S\ref{sec:relv},
the PDR moves relative to the molecular cloud at $-$2.5$\pm$1.1 \kms.
This relative motion is inconsistent with a `stationary' 
pressure-confined nebula where the \HII\ region is confined
within the molecular cloud and the massive
star ionizing the nebula is stationary with respect 
to the molecular cloud. 

\nocite{gf04}Garcia-Segura \& Franco (2004) have considered the
effects due to the natal molecular cloud gravity and stellar motion  
in their recent gas dynamical simulations. These simulations were
done to study the evolution of pressure-confined \HII\ regions. 
They show that a plethora of structures for the ionized gas can 
be produced due to the stellar motion. In such models 
relative motions between the PDR, ionized gas and  
molecular cloud are expected both due to stellar motion
and re-adjustments necessary for the ionized gas to
be in hydrostatic equilibrium in the presence of cloud 
gravity. The PDR may reside in the shock formed
as a result of stellar motion, which also explains 
the high pressure calculated in the PDR compared to the ionized gas.
If the extended \HII\ region W48A is directly associated
with G35.20$-$1.74 (\nocite{kf02}Kurtz \& Franco 2002), 
then, in this picture,  the star has already moved into
a density ramp in the molecular cloud which has resulted 
in the formation of the blister type region 
(\nocite{gf04}Garcia-Segura \& Franco 2004).   
However, other models that are proposed to 
explain the extended lifetime
of \UCHII\ regions cannot be ruled out using the
present data. For example, in the model by 
\nocite{kk01}Kim \& Koo (2001), where they combine
the champagne flow model with the hierarchical structure
of molecular cloud, relative motion of
ionized gas with respect to the molecular cloud
is expected due to the flow. It is possible that the PDR is 
moving with respect to the molecular cloud either
because of stellar motion or due to increased
pressure inside the \HII\ region caused 
by a stellar wind. 
%We await the identification of the
%second line feature to further investigate on the structure
%of G35.20$-$1.74. 

\acknowledgments
We are grateful to the anonymous referee for the critical 
comments and suggestions which have helped in
refining the interpretation of our observations and also
significantly improved the paper. 
DAR thanks S. Sridhar, R. Nityananda, D. Bhattacharya and  
K. S. Dwarakanath for many useful discussions during the course
of the work. We thank C. De Pree and D. Balser for useful comments
and also for proofreading the manuscript. 

\include{tab1}

\include{tab2}

\include{tab3}

\include{tab4}

\include{tab5}

\include{tab6}

\end{document}

%% file: tab1.tex
%Observational parameters

\begin{deluxetable}{lrrrr}
\tabletypesize{\small}
\tablecolumns{4}
\tablewidth{0pc}
\tablecaption{Observational Parameters \label{tab:obs}}
\tablehead{
\colhead{Parameters} & \multicolumn{4}{Values} \\ }
\tablehead{
                     &  6 cm  & 3.6 cm & 2 cm & 0.7 cm }
\startdata
Date of observations       & 21-OCT-2001  & 21-OCT-2001  & 13-OCT-2001 & 13/17-AUG-2004  \\
Field center RA (J2000)    & 19$^h$01$^m$47$^s$.1 & 19$^h$01$^m$47$^s$.1 & 19$^h$01$^m$46$^s$.4                            & 19$^h$01$^m$46$^s$.4  \\
Field center DEC (J2000)   & +01$^{o}$13$^{'}$00$^{''}$ & +01$^{o}$13$^{'}$00$^{''}$ & 
                             +01$^{o}$13$^{'}$24$^{''}$ & +01$^{o}$13$^{'}$24$^{''}$ \\
RLs observed -- IF1        & C110$\alpha$ & C92$\alpha$  & C76$\alpha$, He76$\alpha$ & C53$\alpha$ \\ 
%             -- IF2        & C111$\alpha$ & H92$\alpha$, He92$\alpha$ &  \\ 
             -- IF2        & C111$\alpha$ &         &     &  \\ 
Center freq (GHz) -- IF1   & 4.8754      & 8.3121  & 14.6927 & 42.9638  \\
%                  -- IF2   & 4.74572      & 8.3096  &          \\
                  -- IF2   & 4.7458      &         &       &    \\
Bandwidth (MHz) -- IF1     & 0.78         & 1.56    &  3.13 & 3.13  \\ 
%                -- IF2     & 0.78         & 6.25    &     \\ 
                -- IF2     & 0.78         &     &    &  \\ 
Velocity range (\kms) -- IF1& 48          & 56      &  64  & 42   \\
%                      -- IF2& 49          & 226     &      \\
                      -- IF2& 49          &       &     &  \\
Channel separation (\kms)  &              &       &     &   \\ 
                    -- IF1 & 0.75  & 0.88 & 1.0 & 1.4  \\
%                          -- IF2 & 0.77  & 14.9    &       \\
                    -- IF2 & 0.77  &      &     &      \\
Phase calibrator           & J1851+005 & J1851+005 & J1851+005 &  J1851+005  \\
Bandpass calibrator        & J1229+020 & J1229+020 &  J1924$-$292 & J1733$-$130  \\
                           &           &           &  J1733$-$130 &   \\
Flux calibrator            & 3C 286 & 3C 286 & 3C 286 & 3C 286 \\
On-source observing time (hrs) & 1.6   &  1.8   &  4.0 & 2 $\times$ 3.3  \\
Synthesized beam (arcsec)  & 14.3 $\times$  13.0 & 8.4 $\times$  7.6  & 4.9 $\times$   4.5 & 2.1 $\times$   2.0  \\
Position angle of          &           &          &          &            \\ 
synthesized beam (deg)  & $-$0$^o$.8 & $-$17$^o$.2 & 17$^o$.7 & $-$89$^o$.9  \\
%Synthesized beam (arcsec)  & 14.4 $\times$  13.0 & 8.7 $\times$  8.1 & 5.0 $\times$   4.6  \\
Largest angular size (arcmin) & 5        &    3     &  1.5 &  0.7 \\
RMS noise in the spectral  &           &          &          &           \\
cube (mJy/beam)  & 1.9  & 1.1\tablenotemark{a} & 1.9  & 1.4 \\
RMS noise in the           &           &          &          &           \\ 
continuum images (mJy/beam)  & 2.0  & 1.4 & 2.9 & 0.7   \\
\enddata
\tablenotetext{a}{Spectral RMS obtained from an off source region. The spectra toward
bright continuum emission are affected by baseline ripple and hence have higher RMS.}
\end{deluxetable}

%% file: tab2.tex
%Parameters of the UCHII region G35.20-1.74

\begin{deluxetable}{rcc}
\tabletypesize{\small}
\tablecolumns{3}
\tablewidth{0pc}
\tablecaption{Parameters of G35.20$-$1.74 obtained from the continuum images 
\label{tab:cs}}
\tablehead{
%\colhead{Obs. $\lambda$} & \colhead{Angular Size\tablenotemark{1}} & \colhead{Flux density\tablenotemark{2}} \\
\colhead{Obs. $\lambda$} & \colhead{Angular Size\tablenotemark{1}} & \colhead{Flux density\tablenotemark{2}} \\
\colhead{(cm)}             & \colhead{(\arcsec)} & \colhead{(Jy)} }
\startdata
0.7  & 3.7   & 2.21 (0.02)   \\
%2  & 4.5 $\times$ 4.2 (119$^o$ \tablenotemark{3})   & 2.6 (0.02)   \\
2    & 3.6   & 2.49 (0.02)   \\
%4  & \nodata               & 2.4 (0.02\tablenotemark{2})   \\
4  & $<$ 7.4               & 2.41 (0.02)   \\
%6  & \nodata               & 1.95 (0.03\tablenotemark{2})   \\
6  & $<$ 15.7               & 1.95 (0.03)   \\
\enddata
\tablenotetext{1}{Deconvolved angular size obtained using the AIPS task JMFIT.}
\tablenotetext{2}{The quoted uncertainties are the RMS obtained from the residual 
after removing the continuum model.}
\end{deluxetable}

%% file: tab3.tex
%Physical properties of the UCHII region G35.20-1.74

\begin{deluxetable}{cccccccc}
\tabletypesize{\small}
\tablecolumns{8}
\tablewidth{0pc}
\tablecaption{Physical properties of G35.20$-$1.74 
\label{tab:pp}}
\tablehead{
\colhead{Distance} & \colhead{size} & \colhead{T$_e$} &\colhead{EM ($\times$ 10$^7$)}
&\colhead{n$_e$ ($\times$ 10$^4$)} &\colhead{$U$} &\colhead{Log($N_c$)} &\colhead{Spectral Type\tablenotemark{1}} \\
\colhead{(kpc)}    & \colhead{(pc)} & \colhead{(K)}   &\colhead{(pc cm$^{-6}$)}
&\colhead{(cm$^{-3}$)} & \colhead{(pc cm$^{-2}$)} & \colhead{(s$^{-1}$)}} 
\startdata
%3.2  & 0.1 & 5900 (700) & 4.1 (0.1) & 2.0 & 36.8 & 48.52 & O7.5 \\
3.2  & 0.1 & 9900 (1400) & 6.4 (0.3) & 2.6 & 42.4 & 48.4 & O8 -- O7.5 \\
\enddata
\tablenotetext{1}{Spectral type is estimated by assuming a single star is embedded
in the \UCHII\ region (Panagia 1973).}
\end{deluxetable}

%% file: tab4.tex
%Parameters of the observed RLs

\begin{deluxetable}{rcccccc}
\tabletypesize{\small}
\tablecolumns{7}
\tablewidth{0pc}
\tablecaption{Parameters of the observed recombination lines
\label{tab:line}}
\tablehead{
\colhead{Obs. $\lambda$} & \colhead{Line} & \colhead{$S_L$} & \colhead{$\Delta V $} &\colhead{$V_{LSR}$} & \colhead{$V_{res}\tablenotemark{f}$} & \colhead{Effective Beam\tablenotemark{a}}  \\
\colhead{(cm)}             &                      & \colhead{(mJy)} & \colhead{(\kms)}      &\colhead{(\kms)}    & \colhead{(\kms)}      & \colhead{(\arcsec $\times$ \arcsec)}} 
\startdata
\cutinhead{Toward G35.20$-$1.72\tablenotemark{a}}
0.7     & C53$\alpha$  & 14.7 (5.1) & 5.5(2.2) & 42.0(0.9) & 1.4 & 6.3 $\times$ 6.3  \\
        & X53$\alpha$\tablenotemark{~g}  & 80.0 (6.8) & 3.1(0.3) & 43.4(0.1) & ---   & ---  \\
2       & C76$\alpha$  & 10.7 (1.6) & 4.5(0.8) & 41.9(0.4) & 1.0 & 6.3 $\times$ 6.3  \\
        & He76$\alpha$ & 26.9(0.8)  & 21.2(1.3)& 42.0(0.5) & ---   & --- \\
%2       & C76$\alpha$  & 9.2 (1.8) & 5.3(1.3) & 42.3(0.5) & 1.0 & 5.0 $\times$ 4.6 ($-$0$^o$.34) \\
%        & He76$\alpha$ & 25.0(0.9) & 22.4(1.2) & 42.2(0.5) & - & - \\
3.6     & C92$\alpha$  & (5.2)\tablenotemark{b} &  &  & 0.9 & 8.4 $\times$   7.6 ($-$17$^o$.26) \\
6       & C110$\alpha$ \& & (2.1)\tablenotemark{c} & & & 0.8 & 14.4 $\times$  13.0 ($-$0$^o$.33) \\
        & C111$\alpha$ &  & & & & \\
%\cutinhead{Toward W48A\tablenotemark{d}}
\cutinhead{Toward W48A}
2       & He76$\alpha$ & 1.4 (0.2) & 14.8(1.1) & 47.4(0.5) & 1.0 &  \\
3.6     &              & (0.3) &  & & 0.9 &  \\
6       &              & (1.0)\tablenotemark{e} & & & 0.8 &  \\
\enddata
\tablenotetext{a}{For 0.7 and 2 cm data, the line parameters are obtained 
from the spectra averaged over the effective beam area centered near 
the continuum peak. The coordinates of the continuum peak 
are RA: 19$^h$01$^m$46$^s$.4, DEC: +01$^{o}$13$^{'}$23$^{''}$.9, J2000.} 
\tablenotetext{b}{RMS obtained from the spectrum toward the \UCHII\ region. This
spectrum is affected by a baseline ripple and therefore the estimated RMS is 
higher than that obtained from an off source position.}
\tablenotetext{c}{RMS of the average of C110$\alpha$ and C111$\alpha$ spectra }
%\tablenotetext{d}{see Section~\ref{sec:heline} for details of the spectrum from
%which the line parameters are obtained.}
\tablenotetext{e}{RMS from the C111$\alpha$ spectrum}
\tablenotetext{f}{Channel separation of the spectra in \kms}
\tablenotetext{g}{Presumably a heavy ion (ion massive than C$^+$) recombination line; the LSR velocity is obtained by assuming an infinite mass for the heavy ion.}
\end{deluxetable}

%% file: tab5.tex
%Model parameters of the PDR towards G35.20-1.74

\begin{deluxetable}{rcccc}
\tabletypesize{\small}
\tablecolumns{5}
\tablewidth{0pc}
\tablecaption{Physical properties of the PDR toward G35.20$-$1.74 
\label{tab:pdrp}}
\tablehead{
\colhead{T$_{PDR}$} & \colhead{n$_e^{PDR}$} & \colhead{$l$} 
                               &\colhead{$n_H$} &\colhead{P$_{PDR}$} \\
\colhead{(K)}   & \colhead{(cm$^{-3}$)} & \colhead{($\times$ 10$^{-4}$ pc)} 
&\colhead{($\times$ 10$^{6}$ cm$^{-3}$)} & \colhead{($\times$ 10$^{-7}$ \dycm)}} 
\startdata
\cutinhead{Models with PDR slabs in front and back of the \UCHII\ region}
150  & 1500 -- 2000 & 0.8 -- 0.5  & 5.1 -- 6.8  & 5.3 -- 7.1 \\
200  & 1500 -- 2500 & 1.2 -- 0.6  & 5.1 -- 8.6  & 5.6 -- 9.4 \\
300  & 1500 -- 3000 & 2.6 -- 0.8 & 5.1 -- 10.3  & 6.2 -- 12.5  \\
500  & 1500 -- 4000 & 5.3 -- 1.0 & 5.1 -- 13.7  & 7.5 -- 20.0 \\
1000 & 1500 -- 4000 & 17.0 -- 2.9 & 5.1 -- 13.7 & 10.6 -- 28.4 \\
\cutinhead{Models with a PDR slab in front of the \UCHII\ region}
200  & 2000 -- 2500 & 1.3 -- 1.0  & 6.8 -- 8.6  & 7.5 -- 9.4 \\
300  & 2000 -- 4000 & 2.2 -- 1.0  & 6.8 -- 13.7 & 8.3 -- 16.6  \\
500  & 2000 -- 5500 & 4.5 -- 1.0 & 6.8 -- 18.8  & 10.0 -- 27.5  \\
1000 & 2000 -- 6000 & 14.0 -- 2.6 & 6.8 -- 20.5 & 14.2 -- 42.6 \\
\enddata
%\tablenotetext{a}{The 100 K model is fine tuned to get the line 
%flux at 3.6 and 6 cm wavelengths in absorption 
%within the uncertainty of the observations.}
\end{deluxetable}

%% file: tab6.tex
%VLSR of high density traces observed toward G35.20-1.74

\begin{deluxetable}{llll}
\tabletypesize{\small}
\tablecolumns{4}
\tablewidth{0pc}
\tablecaption{LSR velocity of high-density tracers observed toward G35.20$-$1.74
\label{tab:hdvlsr}}
\tablehead{
\colhead{Molecular} & \colhead{$V_{LSR}$} & \colhead{$\theta_{res}$\tablenotemark{a}} &\colhead{References}  \\ 
\colhead{transition}   & \colhead{(\kms)} & \colhead{(\arcsec)} &  } 
\startdata
\thCO (1 $\rightarrow$ 0)   &  45.0   &  25  & Churchwell \etal (1992) \\
\thCO (2 $\rightarrow$ 1)   &  46.1   &  12  & Churchwell \etal (1992) \\
%\thCO (2 $\rightarrow$ 1)   &  42.4   &  30  & Vallee \& MacLeod (1990) \\
%\CeO (2 $\rightarrow$ 1)    &  42.7   &  30  & Vallee \& MacLeod (1990) \\
%\NHthree (1,1)              &  42.54  &  84  & Anglada \etal (1996) \\
\NHthree (1,1)              &  44.5   &  40  & Churchwell \etal (1990) \\
%\NHthree (1,1)              &  42.5   &  30  & MacDonald \etal (1981) \\
\NHthree (2,2)              &  44.3   &  40  & Churchwell \etal (1990) \\
%CS (1 $\rightarrow$ 0)      &  43.08  &  41  & Anglada \etal (1996) \\
CS (2 $\rightarrow$ 1)      &  42.8   &  25  & Churchwell \etal (1992) \\
CS (5 $\rightarrow$ 4)      &  43.7   &  12  & Churchwell \etal (1992)  \\
%CH$_3$CN(K = 0 \& K = 1)    &  43.9   &  39  & Kalenskii \etal (2000)  \\
\enddata
\tablenotetext{a}{Angular resolution of the observations}
\end{deluxetable}